\documentstyle[prl,aps,psfig]{revtex}
\begin{document}
\draft
%%%%%%%%%%%%%%%%%%
\twocolumn[\hsize\textwidth\columnwidth\hsize\csname
@twocolumnfalse\endcsname
%%%%%%%%%%%%%%%%%%

\widetext
\title{Superconductivity in the two-dimensional $t$-$J$ model} 
\author{S. Sorella,$^{1}$ G.B. Martins,$^{2}$ F. Becca,$^{3}$ C. Gazza,$^{4}$
L. Capriotti,$^{5}$ A. Parola,$^{6}$ and E. Dagotto$^{2}$}
\address{
${^1}$ Istituto Nazionale per la Fisica della Materia, and SISSA, I-34014
Trieste, Italy\\
${^2}$ National High Magnetic Field Lab and Department of Physics, Florida
State University, Tallahassee, Florida 32306 \\
${^3}$ Institut de Physique Th\'eorique, Universit\'e de Lausanne, CH-1015
Lausanne, Switzerland\\
${^4}$ Instituto de Fisica Rosario (CONICET) and Universidad Nacional de
Rosario, Argentina \\
${^5}$ Istituto Nazionale per la Fisica della Materia, Unit\`a di Ricerca
di Firenze,
I-50125 Firenze, Italy \\
${^6}$ Istituto Nazionale per la Fisica della Materia and Dipartimento di
Scienze, Universit\`a dell'Insubria, I-22100 Como, Italy \\
}
\date{\today}
\maketitle
\begin{abstract}
Using computational techniques, it is shown that pairing is a robust 
property of hole 
doped antiferromagnetic (AF) insulators. In one dimension (1D) 
and for two-leg ladder systems, a BCS-like variational wave function with 
long-bond spin-singlets and a Jastrow factor
provides an accurate representation 
of the ground state of the $t$-$J$ model, even though strong quantum
fluctuations 
destroy the off-diagonal superconducting (SC) long-range order in this case.
However, in two dimensions (2D) it is argued -- and numerically confirmed
using several 
techniques, especially quantum Monte Carlo
(QMC) -- that quantum fluctuations are 
not strong enough to suppress superconductivity.
\end{abstract}
]
%%%%%%%%%%%%%%%%%%
%%%%%%%%% PACS USED
%%%%  71.10.Fd Lattice fermion models (Hubbard model, etc.)
%%%%  74.20.Mn Nonconventional mechanisms (spin fluctuations, polarons and
%%%%   bipolarons, resonating valence bond model, anyon mechanism, marginal
%%%%  Fermi liquid, Luttinger liquid,
%%%%  71.10.Pm Fermions in reduced dimensions
%%%%  71.27.+a Strongly correlated electron systems; heavy fermions
%%%%%%%%%%%%%%%%%%
\narrowtext

The nature of high temperature superconductors remains
an important unsolved problem in condensed matter physics.
Strong electronic correlations are widely believed to be 
crucial for the understanding of these materials. Among the several proposed
theories are those where antiferromagnetism induces pairing
in the $d_{x^2-y^2}$ channel~\cite{scalapino}. These approaches
include the following
two classes: (i) theories based on 
Resonant Valence Bond (RVB) wave functions, with electrons paired 
in long spin singlets in all possible arrangements~\cite{anderson,gros}, 
and (ii) theories based on
two-hole $d_{x^2-y^2}$ bound states at infinitesimal
doping, formed to minimize the damage of individual holes 
to the AF order parameter, which condense at finite pair density 
into a superconductor~\cite{review}. 
However, recent
density matrix renormalization group (DMRG) calculations have
seriously questioned these approaches since non-SC striped 
ground states were reported for
realistic couplings and densities in the $t$-$J$ model~\cite{white}. 
Clearly to make progress in the understanding of copper oxides, 
the 2D $t$-$J$ model ground state must be
fully understood, to distinguish among the many proposals. 

In this paper, using a variety of powerful numerical
techniques, the properties of the $t$-$J$ model are
investigated. Our main result is that in the realistic regime of couplings
the 2D $t$-$J$ model supports a $d_{x^2-y^2}$ 
SC ground state, confirming theories of Cu-oxides based
on AF correlations. The $t$-$J$ model used here is 
\begin{equation}\label{tj}
{\cal H} = J \sum_{\langle i,j \rangle} ( {\bf S}_i \cdot {\bf S}_{j} -
\frac{1}{4} n_i n_{j} )
- t \sum_{\langle i,j \rangle \sigma} 
{\tilde c}^\dag_{i,\sigma} {\tilde c}_{j,\sigma} +H.c.,
\end{equation}
where ${\tilde c}_{i,\sigma}$=$c_{i,\sigma}
\left ( 1- n_{i,\bar \sigma} \right )$,
$\langle \dots \rangle$ stands for nearest-neighbor sites, and 
$n_i$ and ${\bf S}_i$ are the electron density and spin at site $i$, 
respectively. Our study focuses on the low hole-doping region of chains, 
two-leg ladders, and square clusters, using different numerical techniques: 
QMC (pure variational and fixed-node (FN) approximations), DMRG, and 
Lanczos. Within our QMC approach, 
it is possible to further improve the variational and FN accuracy 
by applying a few ($p \le 2$) Lanczos steps to the variational ($p=0$) 
wave function $|\Psi_V \rangle$,
$|\Psi_p \rangle= (1+\sum_{k=1}^p \alpha_k H^k )|\Psi_V
\rangle$~\cite{sorella}.
Non-variational estimates of energy and correlation functions can also
be extracted with the variance-extrapolation method~\cite{sorella}.

Our BCS variational wave function is defined as
\begin{equation}\label{wf}
|\Psi_V \rangle = {\cal P}_N {\cal P}_G {\cal J} 
{\rm exp} ( \sum_{i,j}
f_{i,j} c^{\dag}_{i,\uparrow} c^{\dag}_{j,\downarrow} )
|0 \rangle,
\end{equation}
where ${\cal P}_N$ is the projector onto the subspace of $N$ particles,
${\cal P}_G$ is the Gutzwiller projector, which forbids doubly
occupied sites, and ${\cal J} ={\rm exp}(\sum_{i,j} v_{i,j} h_i h_j )$
is a Jastrow factor, defined in terms of the hole density at site $i$, with
$h_i= (1-n_{i\uparrow}) (1-n_{i\downarrow})$, and $v_{i,j}$ being variational
parameters. 
The Fourier transform $f_k$ of the pairing amplitude, $f_{i,j}$,
satisfies~\cite{gros}: 
$f_k = \Delta_k/(\epsilon_k -\mu + \sqrt{(\epsilon_k - \mu)^2 +\Delta_k^2})$,
where $\epsilon_k$ is the free electron dispersion, $\mu$ is the chemical
potential, and $\Delta_k$ is the BCS SC gap function.
In particular, it is known that for square lattices and
periodic boundary conditions (PBC), 
the most relevant contribution to $\Delta_k$ has 
$d_{x^2-y^2}$ symmetry, 
namely $\Delta_k=\Delta (\cos k_x - \cos k_y)$~\cite{gros}. 
For ladders, $\Delta_k=\Delta_x \cos k_x + \Delta_y \cos k_y$ 
was used (the optimized $\Delta_x$ and $\Delta_y$ have
opposite signs), whereas in the 1D case 
$\Delta_k = \Delta_1 \cos  k + \Delta_3 \cos 3 k$.
In the following, VMC denotes results obtained with
$|\Psi_V \rangle$, VMC+$p$LS (with $p=1,2$) those obtained 
with $|\Psi_p \rangle$, and FN and FN+$p$LS results 
obtained using the 
FN approximation with $|\Psi_V \rangle$ and $|\Psi_p \rangle$ as
guiding wave function, respectively. Finally, $0$
variance indicates results obtained with the variance extrapolation method.

The wave function Eq.~(\ref{wf}) describes preformed electron 
pairs, expected to become SC 
within the RVB scenario~\cite{anderson,gros}. An important
component of Eq.(\ref{wf}) is the Gutzwiller projector, 
which at half-filling freezes the charge dynamics, establishing 
quasi-long-range AF order~\cite{caprio}. This shows that
the projected BCS wave function describes magnetic regimes as well.
In addition, the SC order parameter is not simply related to the 
pair amplitude $\Delta$, as in weak-coupling BCS. In fact,
at low hole-doping, it
is proportional to the number of holes, and not
to the number of electrons. This result is natural 
in hole-pairing theories~\cite{review}, where 
superconductivity at half-filling is not possible,
suggesting that such theories maybe similar
to the RVB approach if the latter incorporates long-range singlets, 
and no-double occupancy is enforced. 
Moreover, it was observed that the variational
parameter $\Delta$ decreases with increasing hole doping~\cite{randeria},
suggesting a relation of this quantity with the
pseudogap of underdoped cuprates~\cite{pseudogap}. In
hole-pair based theories, a similar result is obtained with 
the hole binding energy, finite even at half-filling, playing the role of
$\Delta$~\cite{review}.

\begin{figure}
\begin {center}
\vspace{-0.2in}
\mbox{\psfig{figure=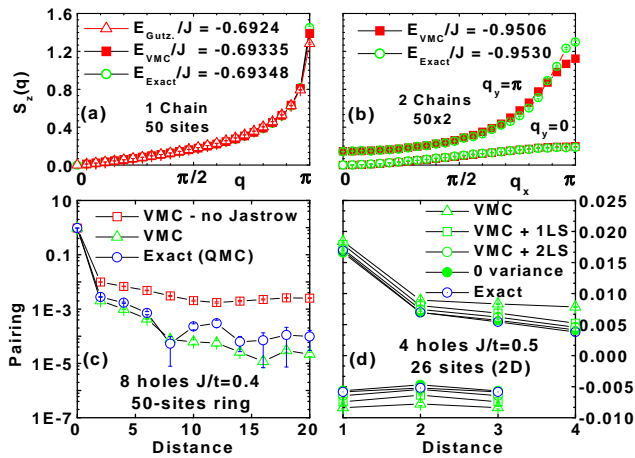,width=3.5in,angle=0}}
\end{center}
\caption{\baselineskip .185in \label{1dpair}
$S_z(q)$ and energies for (a) 1D and (b) 
two-leg ladders at zero hole doping. 
The numerically exact result (open circles) was obtained with the FN method.
Pairing correlations for the $t$-$J$ model in (c) 1D (absolute value) 
and (d) on a tilted 2D cluster for parallel (upper points) and orthogonal
(lower points) singlets.}
\end{figure}

To show that the wave function Eq.~(\ref{wf}) 
accurately describes not just a SC state 
but several magnetic systems as well, 
consider the half-filled model on chains and ladders. 
In the first case, the ground state is 
quasi-antiferromagnetically ordered, with
zero staggered magnetization and power-law spin correlations, 
while in the second case there is a finite spin gap in the 
spectrum, and exponentially decreasing spin
correlations~\cite{ladder}. 
The spin structure factor
$S_z(q)= 1/L \sum_{i,j} e^{i q(R_i-R_j)} \langle S_i^z S_j^z \rangle$ 
shows a cusp at 
$q=\pi$ in 1D, and a broad maximum at $q=(\pi,\pi)$ for two-leg ladders.
These features are remarkably
well reproduced by our variational wave function (Fig.~\ref{1dpair}a-b), which 
generates robust AF correlations at short distances. The state used
is also surprisingly accurate in energy compared to work based 
on a pure Gutzwiller wave function~\cite{gros}, and on numerical
exact studies. It seems that Gutzwiller
projecting the BCS wave function allows for a quantitative
description of AF correlations in low-dimensional 
systems~\cite{comment}. 

Since  undoped systems with short-range AF correlations appear properly
described by the 
wave function Eq.~(\ref{wf}), consider now hole-doped AF systems
where superconductivity should emerge according to some 
theories~\cite{scalapino,anderson,gros,review}.
For this purpose the pairing correlation function 
$\Delta_{i}^{\mu,\nu}(r)$=${\cal S}_{i+r,\mu} {\cal S}^\dag_{i,\nu}$
was studied. Here 
${\cal S}^\dag_{i,\mu}$=
$({\tilde c}^\dag_{i,\uparrow} {\tilde c}^\dag_{i+\mu,\downarrow} -
{\tilde c}^\dag_{i,\downarrow} {\tilde c}^\dag_{i+\mu,\uparrow})$ 
creates an electron singlet pair in the neighboring sites $(i,i+\mu)$. 
Off-diagonal long-range order 
(ODLRO) is implied if $P_d=2 \lim_{r \to \infty}
\sqrt{|\Delta_{i}^{\mu,\nu}(r)}|$
remains finite in the thermodynamic limit.
In 1D, ODLRO is suppressed by quantum fluctuations,
but $\Delta_{i}^{\mu,\nu}(r)$ is finite at short distances and the 
accuracy of the wave function Eq.(\ref{wf}) can be  assessed, the FN
providing exact results in 1D.
As shown in Fig.~\ref{1dpair}c, the 1D pairing correlations are indeed 
non-zero, although rapidly decaying with distance. Our variational 
wave function reproduces accurately the pairing correlations, but
{\it only} when a long-range Jastrow factor is included, otherwise the 
tendency to pairing is overemphasized.
The accuracy of the variational wave function is excellent also 
for small 2D clusters, where the exact solution can be obtained by Lanczos. 
%and the
%0 variance extrapolation of the magnetization at half-filling is
%very close to the exact value~\cite{sorella}.
In this case the quantum fluctuations appear not strong enough to 
destroy superconductivity, see Fig.~\ref{1dpair}d.

\begin{figure}
\begin {center}
\vspace{-0.1in}
\mbox{\psfig{figure=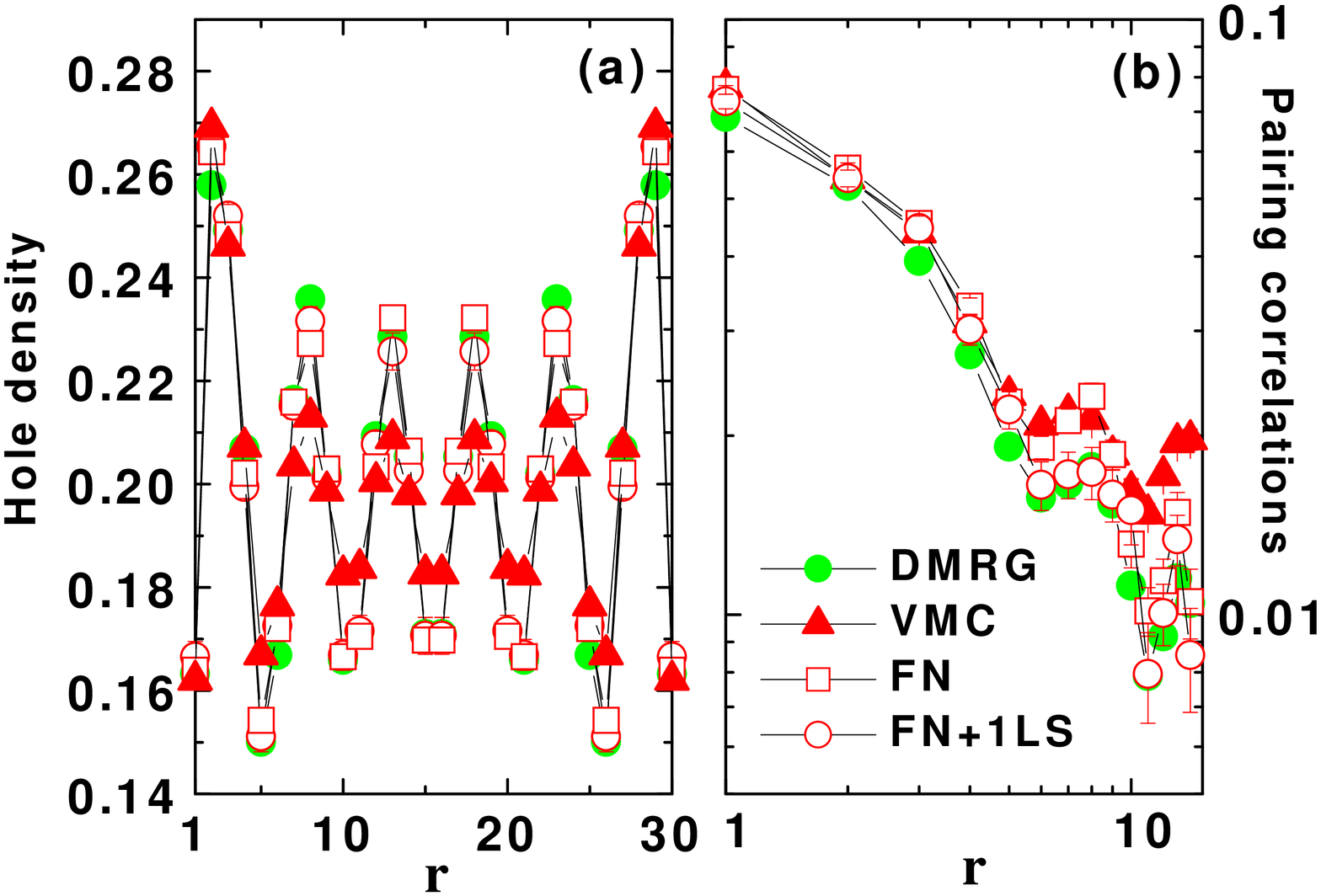,width=3.0in,angle=0}}
\end{center}
\caption{\baselineskip .185in \label{twochains}
(a) Average rung hole density and (b) pairing correlations 
for a $30 \times 2$ ladder, with $12$ holes, $J$=$t$ and 
OBC. In (b) the 
distance $r$ runs from the center of the ladder.}
\end{figure}

Now we consider doped two-leg ladders.
In Fig.~\ref{twochains}, results on a $30 \times 2$ 
ladder with open boundary conditions (OBC) and 
$J$=$t$ are presented~\cite{psladders}. Our
results show that in this case the ground state is qualitatively 
well described by a projected BCS wave function with a 
density Jastrow term, although charge oscillations induced
by hole-pair formation are 
strong in DMRG (here considered as exact), but considerably weaker with VMC.
However, it is remarkable that the FN approach gives the correct rung
density profile, showing that, even in a non-trivial 
system such as a two-leg ladder with hole pairs, 
the correlation functions can be well-controlled by QMC methods.
In general, it is observed that whenever the VMC method 
is not quantitatively accurate, the proper correlation functions 
are obtained by applying the FN approximation~\cite{sandro2}. 
Therefore, this QMC approach combining VMC and FN methods
represents a novel powerful tool 
to assess the reliability of a variational state~\cite{sandro2}, and to obtain
accurate properties of $t$-$J$ models for cuprates~\cite{ceperley}. 

\begin{figure}
\begin {center}
\vspace{-0.2in}
\mbox{\psfig{figure=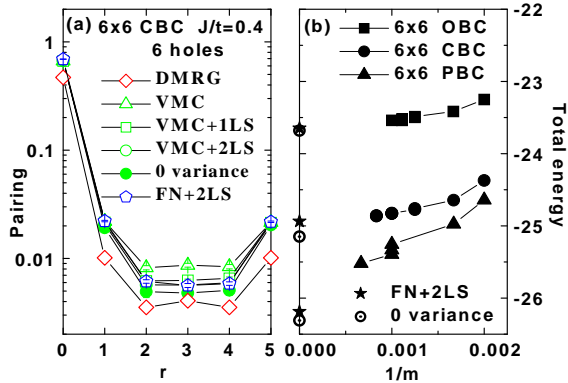,width=3.0in,angle=0}}
\end{center}
\caption{\baselineskip .185in  \label{2dpair}
Pairing correlations (a) and total energy (b) for various 
techniques on 6$\times$6 lattices. $m$ is the number of
states in the DMRG approach. In (a) the distance $r$
runs along the center of the cylinder in the periodic direction.}
\end{figure}

After testing the numerical methods, let us now address the main subject of
the paper, i.e. the possibility of SC in the 2D $t$-$J$ model. For this
case no exact solution (analytical or numerical) is available and,
therefore, it is crucial to perform a careful computational analysis, 
comparing the results of different techniques. While
DMRG allows for an almost exact ground-state characterization
for 1D systems and two-leg ladders, unfortunately
in 2D the results appear to depend on the boundary
conditions~\cite{sandro2}.
In order to compare the performances of QMC and DMRG  
(with $m$ states kept), not only the standard PBC 
have been considered, for which reliable DMRG calculations on large 2D 
clusters presently are not possible, but also the cylindrical boundary 
conditions (CBC), open (periodic) in the $x$ ($y$) direction, 
and OBC in both directions. 
To reduce the number of variational parameters with CBC and OBC, 
the Jastrow factor was restricted to depend only on the distance between
sites, i.e., $v_{i,j}=v_{i-j}$. Site dependent chemical potentials 
$\mu_i$ were also used as additional variational parameters to consider
possible non-uniform charge distributions.
To test the stability of our variational wave function directly
in 2D, a 6$\times$6 lattice with $6$ holes was considered.
In Fig.~\ref{2dpair}, a comparison among different numerical
techniques for several boundary conditions is shown.
The large-$m$ extrapolated energies of DMRG are remarkably similar to those
of QMC for CBC and OBC, but not for PBC where QMC produces substantially
better energies.
As in 1D, for the 6$\times$6 cluster FN does not provide 
qualitative changes in the pairing correlations with respect to the
VMC outcome, and the FN+2LS and VMC+2LS correlations are close. 
In addition, the $m$=1200 DMRG results lead to similar
correlations. The agreement among the many methods
suggests that the QMC method 
correctly reproduces ground state properties also in 2D systems, where
the Jastrow term does not suppress the ODLRO present 
in the VMC, providing sizable long-range pairing.

The most natural
boundaries without  spurious symmetry breaking are 
PBC, since the finite-cluster eigenstates have all the lattice
symmetries.
Therefore, here the subsequent effort building 
toward the main result of the paper focuses on PBC clusters. 
For these boundary conditions, our improved variational calculation 
is accurate and no sign of static stripes has been found at the couplings
investigated, although dynamical effects are still possible
(see below). For $8$ holes on the $8 \times 8$ lattice and
$J/t$=$0.4$, our
best variational energy per site (FN+2LS) is $E$=$-0.66672(6)t$, close
to the zero variance extrapolation $E$=$-0.671(1)t$, and much better than
the pure variational calculation, $E$=$-0.64266(8)t$.

\begin{figure}
\begin {center}
\vspace{-0.1in}
\mbox{\psfig{figure=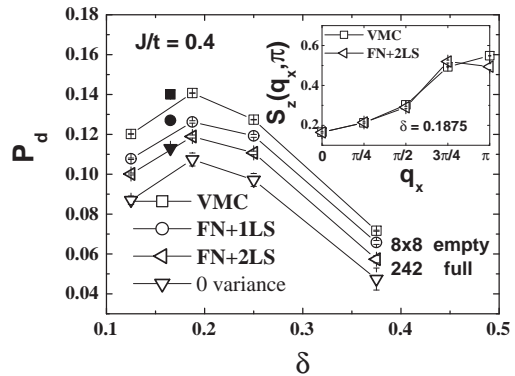,width=3.0in,angle=0}}
\end{center}
\caption{\baselineskip .185in \label{pd}
Superconducting order parameter $P_d$ vs $\delta$ at 
$J/t$=0.4, using the techniques and cluster sizes shown. 
Inset shows $S_z(q)$ at optimal density.}
\end{figure}

The main result of our paper is in Fig.~\ref{pd}, where the SC order 
parameter $P_d$ is shown for clusters of 64 sites
at several densities and 242 sites close to optimal doping. These
results, stable among the many methods used here and with weak size effects,
are indicative of a robust SC ground state away from half-filling in the
$t$-$J$ model.
In view of the success of the present QMC techniques
to reproduce known results for chains, ladders, and OBC 2D clusters, it
is reasonable to consider the
data in Fig.~\ref{pd} as accurate. In addition, the results
are in good qualitative agreement with experimental tendencies,
including an optimal doping at $\delta$$\sim$0.18. The spin 
structure factor in Fig.~\ref{pd} (inset) has 
a broad peak at ($\pi$,$\pi$), concomitant with exponentially decaying
AF correlations in real space. A slight tendency toward
spin incommensurability (SI) appears in the FN+2LS results at optimal density.
At other densities $\delta$, such as
0.25, a similar mild tendency toward SI was also observed.
Then, it is conceivable that the SC state may contain very weak 
dynamical stripe tendencies,
as recently observed in the spin-fermion model~\cite{adriana}. 
Alternatively, band effects or  AF correlations across-holes~\cite{martins}
could be responsible for the SI structure.
In this doping region, the charge structure factor $N(q)$
is found to be basically featureless.

\begin{figure}
\begin {center}
\vspace{-0.2in}
\mbox{\psfig{figure=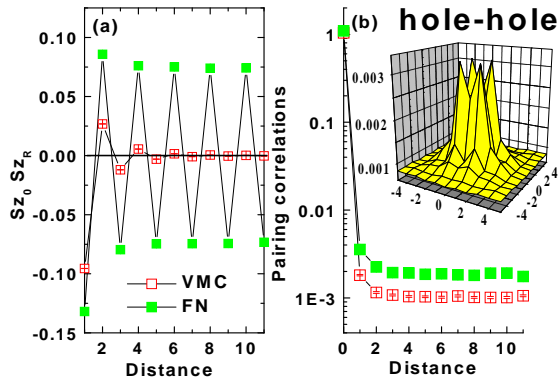,width=3.0in,angle=0}}
\end{center}
\caption{\baselineskip .185in \label{big}
Spin-spin correlations (a) and pairing (b) versus distance. In the inset
of (b) are the hole-hole correlations. The cluster has 242 sites, 8 holes, and 
$J$=$0.4t$.}
\end{figure}

In addition, a low-density of $8$ holes on a tilted 
$242$-site cluster was also analyzed to 
investigate coexistent
antiferromagnetism and superconductivity, as seen in recent experiments for 
underdoped YBCO~\cite{sidis}.
The pure variational approach shows a small SC order
parameter, and a vanishing small AF order 
(Fig.~\ref{big}). Indeed,
at half-filling, the projected $d_{x^2-y^2}$ BCS state underestimates the
magnetic order since it has zero magnetization
with a logarithmically divergent $S_z(\pi,\pi)$~\cite{caprio}, 
and hole doping reduces 
further the AF correlations. Remarkably, the FN approach enhances
{\it both} the SC and AF tendencies. 
In particular, the spin correlations show robust
long-range order implying that antiferromagnetism
survives a small density range~\cite{calandra,chen},
as in experiments~\cite{sidis}. 
The FN energies (not shown) 
for 0,8,24 and 40 holes in 242 sites are clearly stable against phase 
separation, but are quite close to it~\cite{sczhang}.

Moreover, hole-hole correlations display
sharp peaks at hole distance $\sqrt{2}$ 
(see Fig.~\ref{big}), indicating the tendency to have 
short-range hole-pairs in the low hole-density ground state~\cite{review}. 
These pairs were identified in previous studies~\cite{white2}, 
illustrating the good agreement among different techniques.
Using the present QMC method the PBC 6$\times$6 two-holes ground
state was also analyzed.
The results (not shown) once again indicated a clear signal at distance 
$\sqrt{2}$, which always emerges upon Jastrow and/or FN improvement
from an RVB wave function, 
where electrons rather than holes naively 
appear to be paired. This result illustrates the 
flexibility of the QMC approach, and unveils 
unexpected  similarities between the Jastrow corrected
long-singlet RVB~\cite{anderson,gros} 
and the hole-pair approaches~\cite{review}, 
both producing closely related results.

In conclusion, robust indications of superconductivity have been found
in the 2D $t$-$J$ model. The results were
obtained with an RVB wave function with Jastrow factors, flexible enough
to reproduce DMRG results for chains and ladders and
to produce tight hole pairs at low hole density, in agreement with other
techniques. The present effort substantially improves on previous
calculations
where $d$-wave superconductivity was predicted to exist at intermediate
$J/t$~\cite{gros,review},
and highlights the power of the recently developed QMC methods~\cite{sorella}.

This work was partially supported by MURST (COFIN 99), NSF
(DMR-9814350), the NHMFL In-House Research Program,
Fundaci\'on Antorchas, and the PICT grant  N03-03833
(ANPCYT).


\begin{thebibliography}{99}
\vspace{-0.5in}

\bibitem{scalapino} D.J. Scalapino, Phys. Rep. {\bf 250}, 331 (1995).

\bibitem{anderson} P.W. Anderson, Science {\bf 235}, 1196 (1987).

\bibitem{gros} C. Gros {\it et al.}, Z. Phys. B{\bf 68},
425 (1987); 
R. Valenti and C. Gros, Phys. Rev. Lett. {\bf 68}, 2402 (1992).

\bibitem{review} E. Dagotto, Rev. Mod. Phys. {\bf 66}, 763 (1994) and
references therein.

\bibitem{white} S.R. White and D.J. Scalapino, Phys. Rev. Lett. {\bf
80}, 1272 (1998).

\bibitem{sorella} S. Sorella,  Phys. Rev. B{\bf 64}, 024512 (2001);
E. Dagotto and A. Moreo, Phys. Rev. D{\bf 31}, 865 (1985).

\bibitem{caprio} L. Capriotti {\it et al.}, unpublished.

\bibitem{randeria} A. Paramekanti {\it et al.}, cond-mat/0101121.

\bibitem{pseudogap} See for example T. Timusk and B. Statt, Rep. Prog. Phys.
{\bf 62}, 61 (1999) and references therein.

\bibitem{ladder} E. Dagotto and T.M. Rice, Science {\bf 271}, 618 (1996).

\bibitem{comment} 
Eq.(\ref{wf}) is qualitatively different from spin-gapped
short-range RVB states involving only nearest-neighbor singlets.
This is evident from $f_{i,j}$, which is nonzero even at
large distances, allowing for AF correlations.

\bibitem{psladders} For two-leg ladders, 
phase separation occurs at larger $J$'s
[S. Rommer {\it et al.}, Phys. Rev. B{\bf 61}, 13424 (2000)].

\bibitem{sandro2} F. Becca {\it et al.}, to appear in Phys. Rev. Lett.
(2001).

\bibitem{ceperley}
The FN wave function
is not a simple variational state, but is the ground state of a realistic 
FN Hamiltonian [D.F.B. ten Haaf {\it et al.}, Phys. Rev. B{\bf 51},
13039 (1995)].

\bibitem{adriana} Recently, a mixture of dynamical stripes and
superconductivity has been identified in the spin-fermion
model, see A. Moraghebi, S. Yunoki and A. Moreo, preprint.

\bibitem{martins} G.B. Martins {\it et al.},
Phys. Rev. Lett. {\bf 84}, 5844 (2000).

\bibitem{sidis} Y. Sidis {\it et al.}, Phys. Rev. Lett. {\bf 86}, 4100
(2001). See also S. Ono {\it et al.}, Phys. Rev. Lett. {\bf 85}, 638 (2000).

\bibitem{calandra} M. Calandra and S. Sorella, Phys. Rev. B{\bf 61}, 
11894 (2000).

\bibitem{chen} G.J. Chen {\it et al.}, Phys. Rev. B{\bf 42}, 2662 (1990).

\bibitem{sczhang} This result is in agreement with SO(5) 
predictions [S.C. Zhang {\it et al.}, Phys. Rev. B{\bf 60}, 13070 (1999)].

\bibitem{white2}
S.R. White and D.J. Scalapino, Phys. Rev. B{\bf 55}, 6504 (1997).

\end{thebibliography}
\end{document}